\documentstyle[preprint,aps,prl,epsf]{revtex}
\begin{document}
\draft
\preprint{{\bf ETH-TH/96-52}}

\title{Vortices and 2D Bosons: A Path-Integral Monte Carlo Study}
\author{Henrik Nordborg and Gianni Blatter}
\address{Theoretische Physik, ETH-H\"onggerberg, CH-8093
         Z\"urich, Switzerland}
\date{\today}

\maketitle

\tighten

\begin{abstract}
The vortex system in a high-$T_c$ superconductor has been studied
numerically using the mapping to 2D bosons and the path-integral Monte
Carlo method. We find a single first-order transition from an Abrikosov
lattice to an entangled vortex liquid. The transition is characterized
by an entropy jump $\Delta S \approx 0.4 \, k_{\scriptscriptstyle B}$
per vortex and layer (parameters for YBCO) and a Lindemann number $c_L
\approx 0.25$. The increase in density at melting is given by $\Delta
\rho = 6.0\times 10^{-4} / \lambda(T)^2$. The vortex liquid corresponds
to a bosonic superfluid, with $\rho_s = \rho$ even in the limit
$\lambda \rightarrow \infty$.
\end{abstract}

\pacs{PACS numbers: 74.20.De, 74.60.Ec, 05.30.Jp}

Our understanding of the phase diagram of type II superconductors has
improved significantly since the mixed state was introduced by
Abrikosov in 1957\cite{Abrikosov}. The vortex state is of particular
interest for the high-$T_c$ superconductors, where strong thermal
fluctuations lead to melting of the vortex lattice and the appearance
of a vortex liquid phase. Experimental evidence for a first order
vortex lattice melting transition has been obtained from the
observation of jumps in the resistivity and in the
magnetization\cite{Pastoriza,Zeldov,Welp}. More recently, the latent
heat of the transition has been measured directly in an untwinned YBCO
single crystal\cite{Schilling}. Theoretically, vortex lattice melting
has been studied using various approximate techniques, including the
renormalization group, perturbative expansions, density functional
theory, and the Lindemann criterion\cite{Theory}. The absence of one
simple and reliable theory has provoked large interest in numerical
simulations. A number of models, such as the 3D frustrated XY-model,
the lattice London model and the lowest Landau level approximation,
have been used, with no consistent picture emerging,
however\cite{Numerics}. A careful analysis of the 3D vortex system has
been carried out by \v{S}\'{a}\v{s}ik and Stroud \cite{Sasik}, using
the Lowest Landau Level approximation at constant {\em applied} field.
They obtained a jump in the magnetization and could accurately trace
the first-order melting transition in YBCO.

A very fruitful concept was introduced by
Nelson\cite{Nelson}, who showed that the classical statistical
mechanics of the vortex system can be mapped onto the quantum
statistical mechanics of a 2D system of Yukawa bosons, i.~e., bosons
interacting with the potential $V(R) = g^2 {\rm K}_0(R/\lambda)$, where
${\rm K}_0$ is a modified Bessel function, $\lambda$ is the London
penetration depth, and $g^2$ is a coupling constant to be defined
below. We will refer to this system as the Bose model of the vortex
system. A consequence of this mapping is the prediction of a melting
transition into an {\em entangled} vortex liquid, corresponding to a
lattice to superfluid transition in the Bose system.  Unfortunately,
perturbation theory does not work for strongly interacting 2D bosons
and the only quantitative results are from ground-state Monte Carlo
simulations\cite{Magro-Ceperley}, which have not studied the
superfluid density or the excitation spectrum. In this letter, we
present the first path-integral Monte Carlo results for 2D Yukawa
bosons. Both the cases $\lambda$ finite and $\lambda = \infty$ (2D Bose
Coulomb Liquid) are considered. The results are interpreted for the
solid-liquid transition in the vortex system.

In the Feynman path integral formulation, a $d$-dimensional system of
massive quantum particles is equivalent to a classical
$d+1$-dimensional system of interacting elastic strings. The
dimensionless imaginary-time action for $N$ Yukawa bosons is given by
\begin{equation}
{\cal S}/\hbar = \!\int_0^{\beta} \!\! d\tau
\left\{
\sum_i \frac 1
{2\Lambda^2} \left( \frac{d{\bf R}_i}{d \tau} \right)^2 \!\! +
\sum_{i < j} {\rm K}_0 \left( \frac{R_{ij}}{\lambda}\right)
\right\},
\label{action}
\end{equation}
where all energies are measured in units of $g^2$ and all lengths in
units of the particle distance $a_0$ in the crystal, $a_0^2  =
2/\rho\sqrt{3}$, with $\rho$ being the density. The de Boer parameter
$\Lambda$, $\Lambda^2 = \hbar^2 / m a_0^2 g^2$, measures the size of
quantum fluctuations, $\beta = g^2 / T^{\scriptscriptstyle B}$ is the
inverse temperature, and the ${\bf R}_i$ denote the particle positions.
The partition function is the sum over all world lines weighted by this
action and subject to the boundary conditions ${\bf R}_i(\beta) = {\bf
R}_j(0)$, i.~e., every line ends on itself or on some other line. This
action is also valid in the limit $\lambda \rightarrow \infty$ if a
uniform background charge is subtracted. In this case, the Bessel
function reduces to a logarithm.

Redefining parameters, Eq.\ (\ref{action}) can be interpreted as the
free energy for the vortex system in a type II superconductor, ${\cal
S}/ \hbar = {\cal F}/T$\cite{Nelson}, with
\begin{equation}
\Lambda = \frac{T}{a_0\sqrt{2\varepsilon_l \varepsilon_0}}, \quad
\beta = \frac{2\varepsilon_0 L_z}{T},
\label{paramdef}
\end{equation}
where $T$ is the temperature of the vortex system and $L_z$ is the
thickness of the sample. For an anisotropic superconductor, the
elasticity is $\varepsilon_l \approx \varepsilon^2 \varepsilon_0$,
where $\varepsilon^2 = m/M < 1$ is the anisotropy parameter.
Two approximations are required for this mapping from bosons
to vortices: First, the original London functional for the free energy
contains retarded and advanced interactions between the vortex lines,
which are mediated by gauge fields in the Bose picture\cite{review}.
Here, retardation is neglected, equivalent to keeping only the first
term in an expansion around straight lines. Second, it has been argued
that vortex loops in the $ab$-planes are important for vortex lattice
melting\cite{Loops}. A simple estimate for the free energy of a loop of
length $L$, ${\cal F} \approx (L/\xi) (\varepsilon \varepsilon_0 \xi -
T \ln 3)$, shows that loops proliferate in the critical regime close to
$H_{c2}$. We do not consider the critical regime in this work.

\begin{figure}
\epsfxsize=8.0cm
\epsfysize=8.46cm
\centerline{\epsffile{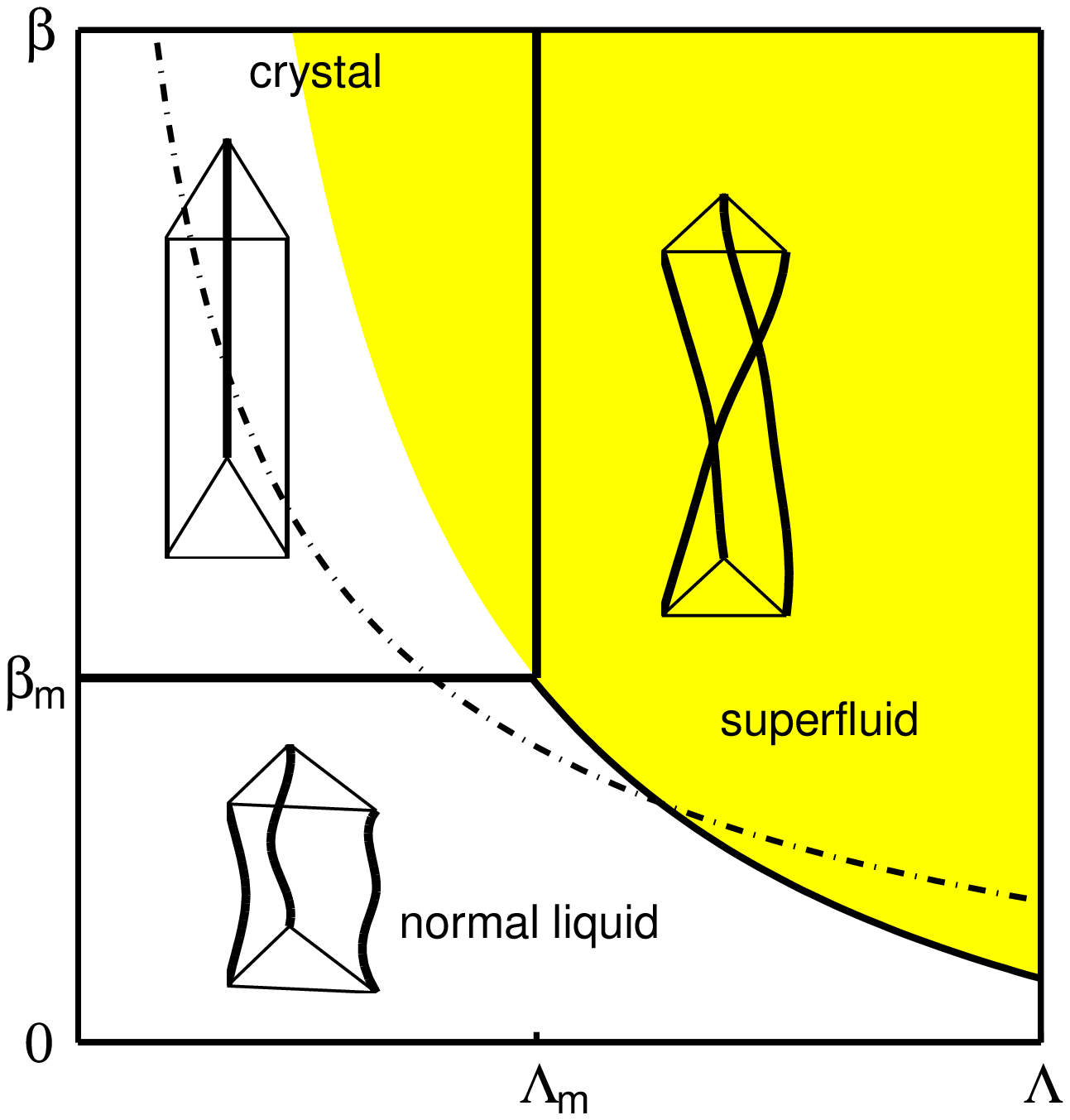}}
\vglue 0.2cm
\caption{Schematic phase diagram for a system of 2D charged bosons in
terms of $\Lambda$ and $\beta$. The solid lines represent phase
transitions and quantum effects are relevant in the shaded region. In
the vortex system, the parameters map to $\Lambda^2 = T^2 /
2\varepsilon_l\varepsilon_0 a_0^2$ and $\beta = 2 \varepsilon_0 L_z /
T$. The  constant field line for a thin sample ($L_z < 70 T /
\varepsilon_0$) is shown (dash-dotted) as it runs through all three
phases. For thicker samples, this line is pushed upwards.}
\label{fig1}
\end{figure}

The phase diagram for the Bose model, shown schematically in
Fig.~\ref{fig1}, contains three phases: A classical high temperature
normal liquid phase, a crystal for low temperatures and small quantum
effects, and a superfluid as quantum effects start to dominate at low
temperatures. This can be understood by considering the three energy
scales involved: The transition from a normal liquid to a
lattice is determined by the competition between the {\em thermal}
energy $T^{\scriptscriptstyle B} = g^2 /\beta$ and the {\em potential}
energy $g^2$. In the limit $\lambda \rightarrow \infty$ and $\Lambda =
0$, the transition takes place at $\beta_m \approx 140$\cite{Caillol}.
With increasing quantum effects, we find a transition from a normal
liquid to a superfluid when the thermal energy matches the {\em
kinetic} energy, $\Lambda^2\beta \approx 1$. At low
temperatures, the competition between potential and kinetic energies
determines whether the system is a crystal or a superfluid. For $\beta,
\lambda = \infty$, it is known that $\Lambda_m =
0.062\cite{Magro-Ceperley}$.

The boson phase diagram can be reinterpreted in terms of the vortex
system, where $\beta$ is proportional to the sample thickness and
$\Lambda$ measures the strength of {\em thermal} fluctuations. For thin
samples, $\beta < \beta_m, 1/\Lambda^2$, we find a {\em disentangled}
vortex liquid. In thicker samples, the system is either a lattice or an
{\em entangled} vortex liquid depending on temperature and magnetic
field. Note the non-trivial mapping between the $H-T$ and
$\Lambda-\beta$ phase diagrams: In a thin sample, the constant field
line (dash-dotted in Fig.~\ref{fig1}) passes through the crystalline
phase (low $T$), the disentangled liquid phase (intermediate $T$), and
the entangled liquid phase (high $T$). With increasing $L_z$, this line
moves to higher values of $\beta$ and the vortex lattice melting line
is determined solely by the value of $\Lambda_m$.

The simulations are carried out using the path integral Monte Carlo
technique, which is exact for bosons\cite{Ceperley:RMP}. Typical runs
involve $N=64$ world lines, though systems as large as $N = 100$ have
been used for analyzing finite-size effects. With $\beta = 300$ we
accuarately capture the ground-state behavior, e.~g., we reproduce the
$T^{\scriptscriptstyle B} = 0$ result of Ref.~\cite{Magro-Ceperley} to
within less than 1\%. We find that $M = 100$ Trotter slices
are sufficient to eliminate systematic errors in the bosonic quantum
phase transition, see Ref.~\cite{Algorithm} for details.
For the largest systems we used roughly 30000 sweeps to equilibrate and
80000 sweeps to measure.

The lattice to liquid transition is
identified through the vanishing of the first Bragg peak (${\bf Q} =
{\bf Q}_1, \tau = 0$) in the structure factor
\begin{equation}
S({\bf Q}, \tau ) = \frac 1 N \left\langle \rho_{\bf Q}(\tau)
\rho_{- \bf Q}(0) \right\rangle,
\label{strucfac}
\end{equation}
where $\rho_{\bf Q}(\tau)$ is the partial Fourier transform of the
density operator, $\rho( {\bf R}, \tau ) = \sum_{i} \delta\left[ {\bf
R} -  {\bf R}_{i}(\tau)  \right]$.
The superfluid density of the Bose system is measured using
the winding number \cite{Ceperley:RMP},
\begin{equation}
\frac{\rho_s}{\rho} = \frac{\langle {\bf W}^2 \rangle}
{ 2\Lambda^2 \beta N},
\end{equation}
with the winding vector ${\bf W} = \sum_i \int_0^{\beta} d\tau \,
\partial_\tau {\bf R}_i$ measuring the diffusion of the center-of-mass
of the system in imaginary time. As the equilibration of the winding
number is very slow for large systems, we use systems with $N = 36$ to
compute the superfluid density. For larger systems, we define the
parameter $\rho_e$ according to
\begin{equation}
\frac{\rho_e}{\rho} = \frac{\mbox{entangled lines}}{\mbox{total number
of lines}},
\end{equation}
where a line is {\em entangled} if it does not end on itself, ${\bf
R}_i(\beta) \neq {\bf R}_i(0)$. The parameter $\rho_e$ measures
the importance of quantum effects in the system, and is therefore
{\it related} to superfluidity. We emphasize, however, that
$\rho_e$ is {\it not} the superfluid density.

We begin by considering the incompressible case with $\lambda =
\infty$. In Fig.~2 we show the results for the first Bragg peak and the
parameter $\rho_e$. The lattice disappears in a sharp transition at
$\Lambda_m \approx 0.062$, in perfect agreement with
ground-state simulations of the same model\cite{Magro-Ceperley}. The
height of the Bragg peak is related to the Lindemann number according
to $S(Q_1) = N \exp( - 8 \pi^2 c_L^2 / 3 )$, and we find $c_L \approx
0.25$. Using Eq.~(\ref{paramdef}), we obtain the melting line
\begin{equation}
B_m(T) =  \frac{4\Lambda_m^2 \Phi_0}{\sqrt{3}}
\frac{\varepsilon_l\varepsilon_0}{T^2},
\label{meltline}
\end{equation}
as expected from a Lindemann criterion with $\Lambda_m \propto
c_L^2$\cite{review}, giving good agreement with experimental
results on YBCO. Its universal character applies only to large magnetic
fields; the finiteness of $\lambda$ leads to a reentrant melting line
at low fields\cite{Nordborg}.

If the vortex lattice melts in a single transition, symmetry requires
it to be first order\cite{Alexander}. In Fig.~\ref{fig3}, we plot the
energy per line and unit length $e \equiv \langle {\cal F} \rangle / N
L_z$. The corresponding energy of the vortex system is defined as
$e_\phi \equiv T^2 \partial_T \ln {\cal Z} / N L_z$, with ${\cal Z}$
the partition function \cite{Hu-MacDonald}. Using
the scaling form  ${\cal F} = \varepsilon_0 a_0 f[\left\{ R_{ij} / a_0
\right\} ]$, valid for large $\lambda$, we obtain $e_\phi = e
(1+t^2)/(1-t^2)$, where $t = T/T_c$. The jump in entropy at the
transition is given by $T _m\Delta s_\phi = \Delta e_\phi$. From Fig.~3
we obtain $\Delta e \approx 0.015 \varepsilon_0$ and therefore
\begin{equation}
\Delta s_\phi [k_{\scriptscriptstyle B}/{\rm length}] \approx 0.03 \,
\varepsilon_0(0) / T_m.
\label{sdef}
\end{equation}
Using parameters for YBCO (layer separation $d = 12$ {\rm \AA},
$\lambda_{ab} \approx 1400 $ {\rm \AA}, and $T_m \approx T_c$) we
obtain $\Delta s_\phi \approx 0.4 k_{\scriptscriptstyle B}$ per vortex
and layer, which compares favorably with the experimental result of
$\Delta s_\phi \approx 0.45\, k_{\scriptscriptstyle B}$
\cite{Schilling}.

For the compressible case with $\lambda$ finite, the statistical
attraction between the bosons produces an increase of the density upon
melting. This maps to a densified vortex liquid due to the
entanglement of the flux lines \cite{Nelson:NAT}. In order to study
this effect, we have developed an isobaric Monte Carlo algorithm which
allows us to fix the external pressure and let the volume
adjust\cite{Algorithm}. The results are shown in Fig.~\ref{fig3} for a
system with $\lambda \approx 1.06 a_0$. The transition is shifted to a
slightly smaller value of $\Lambda_m$ due to the weaker interaction
between the lines. The small shift, less than $4\%$, shows that
Eq.~(\ref{meltline}) gives a good description of the melting for a
large range of fields. The change in density depends on the value of
$\lambda/a_0$; in the present case $\Delta \rho / \rho \approx 0.0003$,
which is consistent with the result (\ref{sdef}) via the
Clausius-Clapeyron relation, $\Delta e / \varepsilon_0 = 8 \pi \Delta
\rho \lambda(T)^2$.

\begin{figure}[t]
\epsfxsize=8.0cm
\epsfysize=7.62cm
\centerline{\epsffile{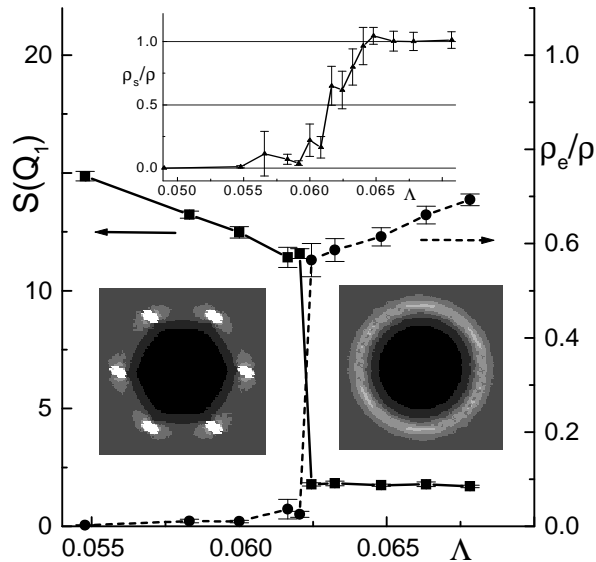}}
\vglue 0.2cm
\caption{The first Bragg peak and the entanglement parameter $\rho_e$
for a system with 64 lines and $\beta = 300$. A sharp transition from a
crystal to an entangled liquid is found at $\Lambda_m = 0.062$. The
structure factors just before and after the melting transition are
displayed. The inset shows the superfluid density for a system with 36
lines.} \label{fig2}
\end{figure}

\begin{figure}
\epsfxsize=8.0cm
\epsfysize=5.74cm
\centerline{\epsffile{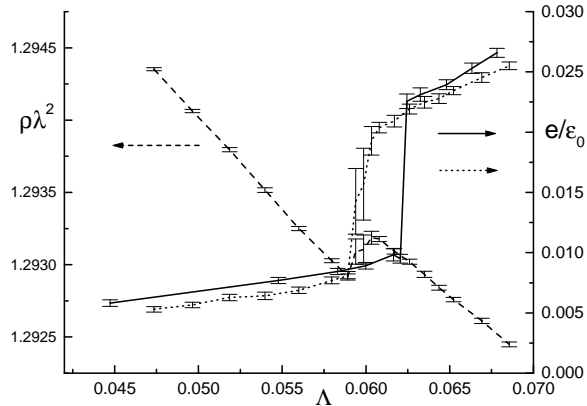}}
\vglue 0.2cm
\caption{The energy $e$ per line of the vortex system (right axis) for
$\lambda = \infty$ (solid line) and $\lambda \approx 1.06 a_0$ (dotted
line). The energy of a perfect lattice with the same density has been
subtracted in both cases. In the compressible system the transition
shows a jump in the density (dashed line). Note the (small)
shift in $\Lambda_m$ as the interaction range
is reduced.}
\label{fig3}
\end{figure}

We turn to the discussion of the liquid phase. In Fig.~\ref{fig2}
we show the entanglement parameter $\rho_e$ which rises sharply at the
transition, indicating that the vortex liquid entangles {\em
immediately} upon melting. This result is in agreement with recent flux
transformer experiments, showing that the vortex correlation along the
magnetic field disappears at the melting transition\cite{Lopez}. The
possibility of a {\em disentangled} vortex liquid has attracted much
interest in recent years\cite{LineLiquid}. Two theoretical arguments
have been put forward in favor of the existence of this phase: To begin
with, the melting of the vortex lattice into an entangled vortex liquid
involves the change of two symmetries: The transverse translational
symmetry of the lattice and the longitudinal gauge symmetry. If these
two symmetries do not change simultaneously, an intermediate phase will
appear. A second argument stems from the analysis of the 2D Bose
Coulomb liquid: It can be shown that the suppression of long wavelength
density fluctuations in this system leads to enhanced phase
fluctuations and an algebraic decay of the off-diagonal long-range
order (ODLRO) even in the ground-state ($T^{\scriptscriptstyle B} =
0$)\cite{Magro-Ceperley}. However, the absence of a
$T^{\scriptscriptstyle B} = 0$ Bose condensate has no straightforward
implication for the superfluid density, which is related to the
excitation spectrum rather than to the ground-state properties. In the
inset of Fig.~2 we show the superfluid density measured by the winding
number. Apart from a slight broadening of the transition due to the
smaller system size, the result shows that $\rho_s = \rho$ as soon as
the translational symmetry is restored in the liquid.
Retardation may
modify this result in the following ways: i) The decrease in the
effective mass of the bosons (elastic tension of the vortices) favors
the entangled state. ii) The retardation may render the entanglement
unstable, thereby favoring a disentangled liquid. Thus, the question
regarding the possibility of a disentangled liquid phase in the
retarded model has not been completely settled.

\begin{figure}
\epsfxsize=8.0cm
\epsfysize=5.77cm
\centerline{\epsffile{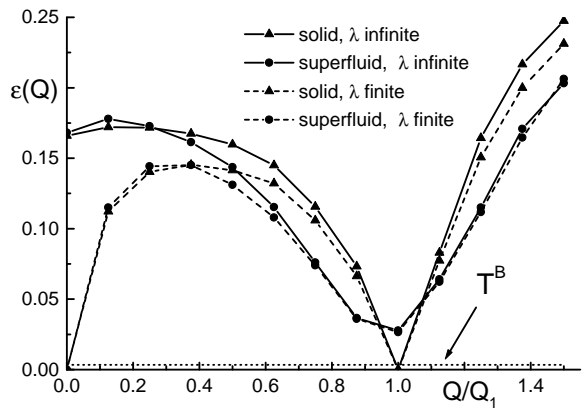}}
\vglue 0.2cm
\caption{The excitation spectra $\varepsilon(Q)$ in units of $g^2$
for a system of 64 lines. Increasing the range $\lambda$ of the
interaction shifts the sound mode to the plasma frequency but leaves
the roton minimum unchanged. The latter collapses upon
crystallization.}
\label{fig4}
\end{figure}

Additional information on the properties of the
Bose superfluid/vortex liquid is provided through the analysis
of the dynamic structure factor. Following Nelson\cite{Nelson},
the partial Fourier transform $S(Q,\tau)$ takes the form
$S(Q,\tau) \approx S(Q,0)\, \exp\{- \varepsilon(Q) |\tau|\}$,
where $\varepsilon(Q)$ is the excitation spectrum of the Bose system.
Thus, the bosonic excitation spectrum defines a longitudinal
correlation length
$l_r = {T}/{2 \Delta_r \varepsilon_0}$ in the vortex fluid, where
$\Delta_r$
denotes the roton minimum. We compute the excitation spectrum
from our simulations by fitting the measured $S(Q,\tau)$
to the single mode approximation,
\begin{equation}
S(Q,i\omega_n) = \frac{C(Q)}{[\omega_n + \Gamma(Q)]^2 +
\varepsilon(Q)^2},
\end{equation}
where $\varepsilon(Q)$ and $\Gamma(Q)$ are the energy and inverse
lifetime of the
excitations. In Fig.~\ref{fig4} we show the resulting spectra both
for the incompressible ($\lambda = \infty$) and the compressible
($\lambda < \infty$) fluids. Most interestingly, the phonon branch
at small $Q$ turns into a plasmon branch as $\lambda \rightarrow
\infty$, while the roton minimum undergoes no visible change. From the
roton minimum $\Delta_r \approx 0.027$ we determine the correlation or
entanglement length at the melting transition, $l_r \approx 1.6 a_0
\sqrt{\varepsilon_l/\varepsilon_0}$, {\it independent} of the
interaction range $\lambda$. Note that in our simulations the roton
gap is much larger than the temperature and we expect to probe
the ground state behavior of the system.

In conclusion, our simulation of the 2D Coulomb Bose model reveals a
single $T^{\scriptscriptstyle B}\approx 0$ quantum phase transition
from a crystal to a
superfluid phase at $\Lambda_m \approx 0.062$. This translates to a
first-order melting transition of the Abrikosov vortex lattice into an
entangled vortex liquid phase. The long-range interaction changes the
bosonic excitation spectrum from phonons to plasmons for small $Q$, but
does not modify the roton minimum. We find that superfluidity is stable
against static long-range interactions.

We are grateful to M. Dodgson, M. Feigel'man, V. Geshkenbein, L. Ioffe,
A. van Otterlo, and H. Tsunetsugu for interesting discussions. One of
us (HN) is particularly grateful to H. Tsunetsugu for ideas on how to
treat the long-range interaction, and to D. Ceperley for valuable
advice on the algorithm. We acknowledge financial support from the
Swiss National Science Foundation.

\end{document}